\documentclass{ws-mpla}

\begin{document}

\markboth{Jennifer Kile}
{Flavored Dark Matter:  A Review}

\catchline{}{}{}{}{}

\title{FLAVORED DARK MATTER:  A REVIEW\\
}

\author{\footnotesize JENNIFER KILE
}

\address{Department of Physics and Astronomy, Northwestern University\\
Evanston, IL 60208,
USA
\\jenkile@northwestern.edu}

\maketitle


\begin{abstract}
The current status of flavored dark matter is reviewed.  We discuss the main experimental constraints on models of flavored dark matter and survey some possible considerations which are relevant for the constructions of models.  We then review the application of existing flavor principles to dark matter, with an emphasis on minimal flavor violation, and discuss implications of flavored dark matter on collider phenomenology.

\end{abstract}

\ccode{PACS Nos.: 95.35.+d, 11.30.Hv}

\section{Introduction}	

The Standard Model (SM) works remarkably well, yet leaves many questions unanswered.  Among these open issues are the solution to the hierarchy problem, the smallness of the cosmological constant, the origin of the baryon-antibaryon asymmetry of the universe, the identity of dark matter (DM), and the origin of fermion flavor and masses.  In formulating models of physics beyond the SM, it is not unreasonable to consider that the addition of new physics may simultaneously address more than one of these questions.  For example, supersymmetry famously solves the hierarchy problem\cite{Dimopoulos:1981yj,Cahn:1981ma,Witten:1981nf} and, as long as R-parity is conserved or nearly conserved, provides a plausible DM candidate\cite{Pagels:1981ke}.  Similarly, models of asymmetric DM\cite{Kaplan:2009ag} have tried to link the observed value of the DM relic density to the baryon asymmetry.  Such strategies often lead to novel and interesting avenues of inquiry.

Here, we review some studies which have attempted to extend the idea of SM flavor to the dark sector.  There are numerous recent examples\cite{Zurek:2008qg,Medvedev:2000pg,Finkbeiner:2007kk,Fan:2013yva} of works investigating scenarios where DM is composed of multiple species, while many past studies have taken DM to be the lightest of a complicated dark sector.  Additionally, there are models\cite{Hashimoto:2011tn,Kajiyama:2011fx} wherein unflavored DM is stabilized or rendered long-lived by requiring DM interactions with the SM fields to obey flavor conservation.  However, in this review, we will focus on scenarios in which DM and the SM fermions both transform under the same flavor group.

Historically, studies of DM have usually assumed it to consist of a single, weakly-interacting particle.  Given the multitude of long-lived species in the visible sector, however, the assumption of a single particle to explain the entirety of DM may be naive.  At the same time, the existence of three generations of fermions has led to many models of high-scale flavor interactions.  As we do not know how DM interacts with itself or with the SM fields, it is reasonable to suppose that perhaps the dark sector contains multiple species and that flavor interactions could be a link between the dark and visible sectors.

Another motivation for flavored DM is the current possible tension between the DM relic density and constraints from direct detection.  As is well-known, the WIMP miracle is the surprisingly good agreement between the weak scale and the scale of DM interactions which would yield the observed DM relic density.  However, the results of direct detection experiments such as XENON100\cite{Aprile:2012nq} place limits of order $\sim 20\mbox{ TeV}$ (for a coupling constant of order unity) on the scale of vector effective four-fermion interactions\cite{Bai:2010hh} between DM and $u$ and $d$ quarks for DM masses between $\sim 20$ GeV and $\sim 1$ TeV; scalar interactions between DM and $u$ and $d$ quarks are even more strongly constrained.  Interactions between the dark sector and leptons or the heavier quarks may help mitigate this tension.

It should be noted that flavored dark matter is a potentially immensely rich subject.  The relatively weak constraints on interactions of DM with leptons\cite{Fox:2011fx} and non-first-generation quarks allow possible implications for collider studies, direct detection, indirect detection, and low-energy flavor observables.  Additionally, there are numerous models of flavor into which DM could be incorporated.  In this review, we attempt to give a rough survey of some possible lines of inquiry and review those results which exist in the literature.

The remainder of this review is organized as follows.  In Sec. \ref{sec:const}, we list some of the most important observables for constraining models of flavored dark matter.  We then give an overview of basic model-building considerations in Sec. \ref{sec:modbuild}.  Sec. \ref{sec:flavmod} discusses the application of existing flavor concepts to DM, with an emphasis on minimal flavor violation.  We then discuss flavored DM collider phenomenology in Sec. \ref{sec:pheno}, and, finally, in Sec. \ref{sec:conc}, we conclude.

\section{Constraining Observables}
\label{sec:const}
Here, we briefly summarize some observables which are likely to be relevant for placing constraints on models of flavored DM.  The observables we discuss here roughly fall into two categories:  constraints from DM properties and constraints from flavor observables.  We will save a discussion of collider phenomenology until Sec. \ref{sec:pheno}.

A basic requirement that any model of DM must fulfill is that it must allow for a relic density in agreement with that observed, $\Omega_{DM}h^2=0.1199\pm 0.0027$\cite{Ade:2013zuv}.  If DM is composed of a single thermal relic, this implies a thermally-averaged annihilation cross-section of $<\sigma |v|>\sim 3\times 10^{-26}\mbox{ cm}^3/\mbox{s}$.  However, flavored DM can introduce a few slight complications into the relic density calculation by virtue of having multiple dark sector states or by having strongly flavor-dependent couplings to SM fields.  

The existence of multiple dark sector states can affect the DM relic density via multiple routes.  For example, it may occur that multiple flavors of DM are stable or sufficiently long-lived to comprise today's relic density.  Alternatively, if the mass splitting between DM species is sufficiently small such that multiple DM species are present at freezeout, coannihilations can be relevant for calculations of the relic density.  Such a case was achieved\cite{Agrawal:2011ze} using an analysis based on minimal flavor violation\cite{Chivukula:1987py,Buras:2000dm,D'Ambrosio:2002ex}; the role of minimal flavor violation in flavored DM models will be expanded on in Sec. \ref{sec:flavmod}.   

Additionally, the interactions between the SM and DM which determine the relic density may be strongly flavor-dependent and vary greatly from model to model.  These interactions may be loop-suppressed\cite{Kumar:2013hfa} or controlled by small Yukawa couplings\cite{Agrawal:2011ze,Batell:2011tc}.  Lastly, we point out that the possibility of achieving the observed relic density via asymmetric DM has briefly been considered\cite{Kumar:2013hfa}.

A second basic requirement of DM candidates is that they evade current constraints from direct detection.  Current results from XENON100\cite{Aprile:2012nq} place upper bounds on the spin-independent DM-nucleon cross-section as low as $\sim 2\times 10^{-45}\mbox{ cm}^2$.  Like the relic density, direct detection cross-sections in flavored DM models can have some interesting features.  Among these features is that the interactions of flavored DM with quarks can have important dependence on Yukawa couplings and CKM angles; for example, in some scenarios\cite{Batell:2011tc}, the DM direct-detection cross-section can be dominated by interactions via the $b$ quark.  Alternatively, the DM-nucleon cross-section may be loop-suppressed, such as in the case of DM that interacts preferentially with leptons or third-generation quarks\cite{Agrawal:2011ze}.  However, more exotic direct detection signatures are also possible.  If splittings between DM flavors are sufficiently small that multiple species comprise DM today, exothermic down-scattering in direct detection experiments is possible\cite{Kile:2011mn}; in such a scenario, an incoming DM particle $\chi'$ scatters into a lighter flavor $\chi$, giving a final state with more kinetic energy than what would occur in elastic scattering.  Conversely, up-scattering can occur, rendering flavored DM an example of inelastic dark matter\cite{TuckerSmith:2001hy}; larger mass splittings may effectively eliminate the spin-independent DM-nucleon cross-section; this can be achieved for flavored DM by, for example, using small Majorana mass terms\cite{Kumar:2013hfa}.

We now discuss the relevance of flavor observables.  If the DM interacts with quarks, significant constraints may be obtained from meson mixing.  Lower bounds on the new physics scale for tree-level contributions to  $K^0-\bar{K}^0$ mixing are ${\cal O}(10^3-10^4\mbox{ TeV})$ for CP-conserving interactions; constraints are tighter in the case of CP violation\cite{Isidori:2010kg}.  Dark sector fields can contribute to $K^0-\bar{K}^0$ mixing via loop diagrams which may be additionally suppressed by small Yukawas or CKM angles\cite{Batell:2011tc,Agrawal:2011ze}; in specific flavor models, new physics contributions of loops of visible-sector particles may also have to be taken into account\cite{Kile:2011mn}.  Constraints can similarly be obtained from $B^0-\bar{B}^0$ and $D^0-\bar{D}^0$ mixing.

In addition to meson mixing, rare decays could also be enhanced via contributions from loops of dark sector particles.  For quark-flavored DM, the decay $b\rightarrow s\gamma$ may be relevant; similarly, $\mu\rightarrow e\gamma$ may give useful constraints on leptonically-interacting flavored DM.

If flavored DM is sufficiently light, it can also show up as missing energy in particle decays.  Constraints on effective operators coupling light DM to SM fields can be derived from the flavor-changing decays $s\rightarrow dX$, $b\rightarrow sX$, and $b\rightarrow dX$, where $X$ is an invisible (possibly multiparticle) state; such constraints are strongly dependent on the interaction considered\cite{Kamenik:2011vy}.  Additionally, decays of the top quark, $t\rightarrow j+X$, can be relevant for models of flavored DM\cite{Kamenik:2011nb}; $5\sigma$ evidence for such a decay is expected at the 14 TeV LHC for branching fractions greater than $7\times 10^{-5}$, assuming $100 \mbox{ fb}^{-1}$ of integrated luminosity\cite{Li:2011ja}.

These observables will play a part in specific constructions of flavored DM, to which we now turn.

\section{Model-building considerations}
\label{sec:modbuild}

As the intersection of DM and flavor can yield a potentially immense space of ideas to explore, here we schematically outline some of the choices one may make when building a model of flavored DM.  In doing so, we will point out avenues which may be investigated and discuss those scenarios which currently exist in the literature.

In addition to basic DM properties such as its mass and spin, flavored DM carries flavor quantum numbers which must be specified.  This requires, however, also specifying the flavor quantum numbers of the SM fermions.  For example, one could devise models where all of the SM fermions carry flavor quantum numbers.  Alternatively, one may speculate that the top quark, having a mass very close to the electroweak scale, may have interactions not shared with the leptons or other quarks.  Additionally, one may consider flavor models where the only SM fields that participate are the charged leptons or the neutrinos.  Finally, for each one of these options, we can also choose to have the flavor symmetry act on just the left-handed fields, just the right-handed fields, or both.

Related questions concern the symmetry group itself.  Is the symmetry global or local?  Is the DM in the same representation of the symmetry group as some subset of the SM fermions?  If the symmetry is global, what mediates interactions between the dark sector and the SM?  Are the interactions between the SM and the dark sector renormalizeable or represented by higher-dimension operators?  Some possible answers to each of these questions will be explored as we discuss specific models below. 

We now review some models currently in the literature.  We will group these models into four categories by the type of flavor they carry.  The first of these categories, quark-flavored DM, is the most extensively studied; we will initially focus on interactions which involve the light (i.e., non-top) quarks.  Next, we look at an example of top-flavored DM, with its clear relevance to collider phenomenology.  Then, we consider charged-lepton-flavored DM, and, lastly, neutrino-flavored DM.

Quark-flavored DM has been studied in several works.  The SM quarks obey an approximate global $SU(3)_Q\times SU(3)_u\times SU(3)_d$ symmetry broken by the quark Yukawa couplings, where $Q$, $u$, and $d$ are the left-handed quark $SU(2)_{EW}$ doublet and the right-handed $SU(2)_{EW}$ up- and down-type singlets, respectively.  A model\cite{Batell:2011tc} based on the principle of minimal flavor violation considered scalar DM charged under the same flavor group as the left-handed quark doublets, $SU(3)_Q$.  This scalar then couples to the SM quarks (and possibly the Higgs, which is kept as a flavor singlet in this model) through effective operators which respect minimal flavor violation.  The consequences of minimal flavor violation in this model will be discussed in more detail in Sec. \ref{sec:flavmod}. 

Interactions between DM and the right-handed quarks have also been studied, again within the framework of minimal flavor violation\cite{Agrawal:2011ze}.  In this model the DM is a Dirac fermion which couples to the right-handed quarks via renormalizeable operators which include a new charged scalar.  Three separate cases where the DM transforms under $SU(3)_u$, $SU(3)_d$, and $SU(3)_Q$ are considered.  The role of minimal flavor violation in this model will also be discussed in Sec. \ref{sec:flavmod}.

It is also possible to build models where not all of the SM quarks transform under the flavor symmetry.  DM which interacts specifically with $d$ and $s$ quarks has been investigated\cite{Kile:2011mn}; in this case vector and purely right-handed interactions between DM and the light quarks were considered, with the idea of possible applicability to gauged flavor models.  Toy models based on a gauged $SU(2)_F$ flavor symmetry were constructed, where two dark fields transformed as a doublet under the $SU(2)_F$; the $d$ and $s$ (or, in the case of vector interactions, the light quark doublets) transformed similarly.  These models were anomaly-free and produced no tree-level contribution to $K^0-\bar{K}^0$ mixing.

Flavored DM which interacts with both the light quarks and the top quark can be especially interesting with regard to collider signatures.  Interactions of DM with quarks leading to monotop\cite{Andrea:2011ws} signatures have also been considered\cite{Kamenik:2011nb}.  These interactions are parameterized in terms of dimension-six operators which couple a pair of DM fields to a pair of quarks, one of which is a top quark; vector and scalar operators were considered.  The relevant flavor models and collider signatures will be discussed in Secs. \ref{sec:flavmod} and \ref{sec:pheno}, respectively.

Dark matter which specifically carries top flavor was investigated\cite{Kumar:2013hfa} as a possible explanation for the top forward-backward asymmetry\cite{Aaltonen:2012it,Abazov:2011rq}.  In this model, the DM, $\chi_t$, interacts with the top quark via a scalar mediator $\phi$ with couplings which obey minimal flavor violation.  The $\phi$ is pair-produced with a forward-backward asymmetry via a $t$-channel process and then decays to $t\bar{\chi}_t$, thus passing this forward-backward asymmetry on to the top quark.  The relic density and the cross-section of the $\chi_t$ with nucleons relevant for direct detection are controlled by interactions induced at 1-loop order.  The stringent spin-independent direct detection limits are evaded in this model by giving $\chi_t$ a small Majorana mass term, splitting it into two states and rendering its spin-independent interaction inelastic; for a sufficiently large mass splitting, spin-independent scattering becomes kinematically inaccessible.

In comparison with quark-flavored DM, DM with the flavor quantum numbers of the charged SM leptons is, as of yet, relatively unexplored.  Renormalizeable interactions between flavored DM and the right-handed charged leptons $e$, $\mu$ and $\tau$ analogous to those with $u$ and $d$ above have been investigated\cite{Agrawal:2011ze}.  Similar to the quark case, the DM is a Dirac fermion which transforms under $SU(3)_e$ or $SU(3)_L$, where $e$ and $L$ denote the right-handed charged leptons and the lepton $SU(2)_{EW}$ doublet, respectively.  As before, these interactions are mediated by a charged scalar.  Unlike the case of quark-flavored DM, DM which interacts only with leptons necessarily has loop-suppressed cross-sections for interactions with nucleons and, thus, potentially looser constraints from direct detection.

Finally, we briefly mention connections between neutrino flavor and DM.  Several works have linked DM to neutrinos using discrete symmetries\cite{Meloni:2011cc,Kajiyama:2010sb,Ahn:2012cga,Barry:2011fp,Eby:2011qa}.  Also, we note that sneutrinos can provide a neutrino-flavored DM candidate\cite{MarchRussell:2009aq,Kumar:2009sf}.  Lastly, there are many models of sterile neutrino DM, such as the $\nu$MSM\cite{Asaka:2005an} where the identity of the DM state is closely related to masses and mixings of the active SM neutrinos.

\section{Application of flavor models to DM}
\label{sec:flavmod}
Here we discuss a few principles of flavor into which DM has been successfully incorporated.  We will concentrate on minimal flavor violation, but will also touch upon a few other features of flavor models which have proven useful in connection to DM.

\subsection{Minimal flavor violation}

Several of the flavored DM works mentioned above have utilized minimal flavor violation.  The first of these used minimal flavor violation to stabilize the DM\cite{Batell:2011tc}.  Treating the quark Yukawa matrices $Y_u$ and $Y_d$ as spurion fields, they must transform, respectively, as $(3,\bar{3},1)$ and $(3,1,\bar{3})$ under the global flavor symmetry $SU(3)_Q\times SU(3)_u \times SU(3)_d$.  Their DM candidate, $\chi$, can then potentially decay to SM fields through operators of the form
\begin{equation}
{\cal O}_{decay}=\chi \underbrace{Q \ldots}_{A}\underbrace{\bar{Q} \ldots}_{B}\underbrace{u \ldots}_{C}\underbrace{\bar{u} \ldots}_{D}\underbrace{d \ldots}_{E}\underbrace{\bar{d} \ldots}_{F} \underbrace{Y_u \ldots}_{G}\underbrace{Y_u^{\dagger} \ldots}_{H} \underbrace{Y_d \ldots}_{I}\underbrace{Y_d^{\dagger} \ldots}_{J} {\cal O}_{weak},
\end{equation}
where $A$, $B$, etc. are the numbers of the respective $Q$, $\bar{Q}$, etc., and ${\cal O}_{weak}$ is a possible electroweak operator which does not carry quark flavor.  ${\cal O}_{decay}$ must be a singlet under color; as $\chi$ is taken to be colorless, this implies
\begin{equation}
(A-B+C-D+E-F)\mbox{ mod }3=0.
\end{equation}
Similar relations hold for the flavor $SU(3)$ symmetries.  Taking $\chi$ to transform as 
\begin{equation}
\chi\sim (n_Q,m_Q)_Q \times (n_u,m_u)_u \times (n_d,m_d)_d,
\end{equation}
where the $n_i$ and $m_i$ are integers that indicate the number of $3$ and $\bar{3}$ factors, respectively, these relations are 
\begin{align}
(n_Q-m_Q+A-B+G-H+I-J)\mbox{ mod }3&=0,\nonumber\\
(n_u-m_u+C-D-G+H)\mbox{ mod }3&=0,\\
(n_d-m_d+E-F-I+J)\mbox{ mod }3&=0.\nonumber
\end{align} 
From these equations, it follows that 
\begin{equation}
(n_Q+n_u+n_d-m_Q-m_u-m_d)\mbox{ mod }3=0.
\end{equation}
If this relation does not hold, ${\cal O}_{decay}$ cannot be constructed, and the DM is rendered stable without the imposition of additional discrete symmetries.

The principle of minimal flavor violation can also constrain the mass spectrum of dark states or their interactions with the SM fields.  In the same work\cite{Batell:2011tc}, the authors investigate scalar DM which transforms as an $SU(3)_Q$ triplet $S$; the mass terms are taken to obey minimal flavor violation,
\begin{equation}
-{\cal L}_{mass} =  S^{\dagger}(m_A^2 + m_B^2 Y_u Y_u^{\dagger} + \ldots)S
\end{equation}
where insertions of $Y_d$ have been neglected.  Due to the Yukawa matrices multiplying $m_B^2$, the masses of the three $S$ states will split, such that $m_1\approx m_2$, but $m_3$ will be larger or smaller, depending on the sign of $m_B^2$.  For the case of an ``inverted'' mass spectrum, $m_3<m_1, m_2$, and $S_3$ comprises DM.  They then concentrate on an interaction between $S$ and the SM of the form
\begin{equation}
{\cal L}_{eff} = \frac{c}{\Lambda^2}(\bar{Q}S)(S^{\dagger}Y_dd)H,
\end{equation}
where the flavor indices are contracted within the parentheses.  This interaction, through which the DM $S_3$ interacts predominantly with $b$ quarks, then determines the relic density and DM-nucleon cross-section relevant for direct detection.

Minimal flavor violation has similarly been used in models where DM couples to the right-handed quarks through renormalizeable operators\cite{Agrawal:2011ze}.  Depending on whether the DM transforms under $SU(3)_Q$, $SU(3)_u$, or $SU(3)_d$ and whether it couples to up-type or down-type quarks, the couplings of the DM to the SM can be either similar or hierarchical due to the insertions of Yukawa matrices needed to make flavor singlet interactions.   The same authors applied the same strategy to leptonic-flavored DM; they considered DM which transforms under $SU(3)_L\times SU(3)_e$, where $L$ and $e$ represent the lepton $SU(2)_{EW}$ doublets and charged singlets, respectively.  They take an interaction where fermionic DM $\chi$ interacts with the right-handed singlets as
\begin{equation}
\lambda\bar{\chi}\phi^{\dagger}\ell, 
\end{equation}
where $\lambda$ is a coupling, $\phi$ is a scalar charged under the SM and $\ell$ is a flavor $SU(3)_e$ triplet of the right-handed charged leptons.  Like the above cases, the mass spectrum will typically consist of two roughly degenerate states, $\chi_e$ and $\chi_{\mu}$, and one state, $\chi_{\tau}$ which differs in mass significantly from the other two.  The form of $\lambda$ allowed by minimal flavor violation depends on whether the DM transforms under $SU(3)_L$ or $SU(3)_e$; for $\chi$ which transforms as $SU(3)_e$, $\lambda$ may take the form
\begin{equation}
\lambda= (a+b Y_e^{\dagger} Y_e),
\end{equation}
where $Y_e$ is the charged lepton Yukawa matrix and $a$ and $b$ are constants.  For $SU(3)_L$ DM, the $\lambda$ must be proportional to the Yukawa coupling,
\begin{equation}
\lambda=c Y_e.
\end{equation}
This latter option gives couplings of the $\chi$ states to the SM which are strongly hierarchical.  Thus, in this latter scenario, the DM-SM interactsions will strongly depend on whether $\chi_e$ or $\chi_{\tau}$ comprises DM. 

In the case of top-flavored DM\cite{Kumar:2013hfa}, the authors used minimal flavor violation to make their DM candidate couple preferentially to top quarks.  Their fermionic DM flavor multiplet $\chi$ transforms as $(1,3,1)$ under $SU(3)_Q\times SU(3)_u \times SU(3)_d$; its mass terms can be written as
\begin{equation}
\bar{\chi}(m_0+m_1Y^{\dagger}_uY_u+\ldots)\chi = m_{\chi_u}\bar{\chi}_u\chi_u + m_{\chi_c}\bar{\chi}_c\chi_c + m_{\chi_t}\bar{\chi}_t\chi_t,
\end{equation}
where the down-type Yukawa couplings have been neglected.  As the $u$ and $c$ Yukawa couplings are much smaller than that of the $t$ quark, $\chi_u$ and $\chi_c$ can be roughly degenerate; for $m_0$ positive, $m_1$ negative, $\chi_t$ can be lighter than $\chi_u$ and $\chi_c$ and thus function as the DM candidate.  They then couple their DM candidate to the SM quarks via an unflavored scalar mediator, $\phi$, via operators of the form
\begin{equation}
\bar{q}(g_0 +g_1 Y^{\dagger}_uY_u+\ldots)\chi\phi = g_u\bar{u}\chi_u\phi + g_c\bar{c}\chi_c\phi + g_t\bar{t}\chi_t\phi,
\end{equation}
where $q$ is a flavor triplet of right-handed quarks, $(u, c, t)$.  Similar to the mass terms, $g_u\approx g_c\approx g_0$, while $g_t$ may be somewhat different.  The coupling of the DM candidate $\chi_t$ only to $t$ quarks is then used to address the $t\bar{t}$ asymmetry.  Minimal flavor violation has also been used with higher-dimensional operators to enhance the ratio of the monotop to monojet signal, possibly observable at LHC\cite{Kamenik:2011nb}.

\subsection{Other flavor principles applied to DM}
We briefly mention a few other principles common in flavor studies which have been applied to flavored DM.  Among these are Froggatt-Nielsen models\cite{Froggatt:1978nt}, which have been briefly considered within the context of monotop production\cite{Kamenik:2011nb}; for proper assignment of Froggatt-Nielsen charges, vertices such as $\bar{t}u\chi^2$, where $\chi$ is DM, appear to violate, but in fact conserve flavor.  Of course, other applications of Froggatt-Nielsen models are in principle possible.  Additionally, some models exploiting discrete symmetries\cite{Ahn:2012cga,Barry:2011fp,Hirsch:2010ru} to link neutrinos and DM are based on $A_4$ models which have been used in the neutrino sector to achieve tri-bimaximal mixing\cite{Ma:2004zv}.

\section{Collider phenomenology}
\label{sec:pheno}
Collider signatures of flavored DM vary greatly between models.  Here we mention some of those which have appeared in the literature. 

In the case of models where flavor is gauged\cite{Kile:2011mn}, heavy flavor gauge bosons can potentially be produced in colliders.  If these gauge bosons decay to DM particles, they may produce monojet\cite{Bai:2010hh,Rizzo:2008fp}, mono-$Z$\cite{Petriello:2008pu,Carpenter:2012rg}, or mono-photon\cite{Gershtein:2008bf,Fox:2011pm} signatures.  (These signatures can also be considered simultaneously\cite{Zhou:2013fla}.)  Such gauge bosons could decay to the SM fermions which are also charged under the flavor group.  Additionally, it is possible that such flavor gauge bosons may decay to unstable dark sector particles, which then decay in the detector to partially visible states.

Top-flavored DM has been studied within the context of the Tevatron $t\bar{t}$ asymmetry\cite{Kumar:2013hfa}; in this scenario, a scalar $\phi$ with SM quantum numbers of a right-handed top quark is pair-produced at the Tevatron with a significant forward-backward asymmetry.  $\phi$ then decays to a top quark and top-flavored DM $\chi_t$, communicating the forward-backward asymmetry to the observed top quark-antiquark pair; the model can also produce a top charge asymmetry at LHC.  Additionally, the DM $\chi_t$ is one member of a flavor triplet, along with $\chi_u$ and $\chi_c$; the heavier $\chi_u$ can also be produced, decaying to $u\bar{t}\chi_t$, which can be relevant for top-jet resonance searches.

Collider signatures of $\tau$-flavored DM have also been investigated\cite{Agrawal:2011ze}.  In this model, a charged scalar $\phi$ decays down to leptons and the DM $\chi_{\tau}$, possibly through a cascade of charged leptons and dark sector particles $\chi_e$ and $\chi_{\mu}$.  The final state is expected to contain exactly two $\tau$ leptons, missing energy, and additional leptons; flavor correlations may be used to distinguish between $\tau$-flavored DM and other models.

We briefly mention a few other collider signatures of flavored DM.  The possibility of Higgs decays to flavored dark states has also been considered\cite{Batell:2011tc}.  Study of the monotop signature\cite{Kamenik:2011nb} also appear relevant for LHC.  Additionally, cascade decays similar to those for $\tau$-flavored DM above could also occur for quark-flavored DM, giving final states consisting of missing energy and heavy flavor quarks.  Also, we emphasize that the experimental signatures presented here are limited to those already explored in the literature; many other collider signatures are presumably possible.

\section{Conclusions}
\label{sec:conc}
Both flavor physics and DM are active areas of investigation which necessarily require physics beyond the SM, and the possibility that they could be related in some way is a highly intriguing idea.  The hypothesis that DM carries flavor quantum numbers and interacts with the SM in a flavor-dependent way opens many avenues of inquiry.  Flavored DM is sensitive to constraints from DM observables and precision flavor measurements and potentially provides a wealth of possible experimental signatures.

Many possible lines of investigation into flavored DM exist, a few of which have been described briefly here.  Those models currently in the literature have mainly concentrated on interactions between DM and the SM quarks.  However, they have differed significantly in their motivation, DM flavor structure, the form of the DM-SM interactions, and their collider phenomenology.

We wish to stress, however, that the intersection of flavor physics and DM is an immensely rich subject which still remains relatively unexplored.  Numerous possibilities exist with respect to the choice of flavor symmetry and which SM fields participate in the interaction.  Leptonically-flavored DM, in particular, is ripe for further investigation.  Additionally, one can attempt to incorporate DM into existing models of flavor.  In conclusion, flavored DM has significant potential for future exploration.

\section{Acknowledgements}
The author would like to thank A. de Gouv\^{e}a and W.-C. Huang for helpful comments on the draft.  This work was supported in part under US DOE contract No. DE-FG02-91ER40684.

\end{document}